\documentstyle[12pt]{article}
\vspace{5mm}
\title{A NEW APPROACH TO PERTURBATIVE THEORY IN THE NONPERTURBATIVE
QCD  VACUUM.}
\vspace{2mm}
\author{D.V.Antonov \\
Institute of Theoretical and
Experimental Physics, \\
B.Cheremushkinskaya,25, 117218, Moscow, Russia}
\date{}

\begin{document}
\maketitle

\newcommand{\be}{\begin{equation}}
\newcommand{\ee}{\end{equation}}

\vspace{10mm}

\large

\centerline{\bf Abstract}

\vspace{5mm}
Using stochastic quantization method [1], we derive
gauge--invariant equations, connecting multilocal vacuum correlators of
nonperturbative field configurations, immersed into the quantum background.
Three alternative methods of stochastic regularization of these equations
are suggested, and the corresponding  regularized propagators of a
background field are obtained in the lowest order of perturbation theory.

\vspace{10mm}
\section{Introduction.}

\vspace{3mm}
\hspace{5mm} Recently a new approach to investigation of the nonperturbative
content of any field theory, based on the equations for vacuum correlators,
derived via stochastic quantization method [1], was suggested [2]. It is
closely connected to the Method of Vacuum Correlators [3], in which it is
postulated, that the whole information about the QCD vacuum structure is
maintained in the full set of irreducible vacuum averages (cumulants).

Within this approach, applied to gauge theories, introducing correspon--
ding generating functionals and using cumulant expansion [4,5], one obtains
an infinite set of equations. These equations connect correlators, which
contain various number of fields and stochastic Gaussian noise in
gauge-invariant way. This approach is especially useful in the theories
with nontrivial vacuum structure, e.g. in QCD, since the asymptotics of
solutions of these equations  at Langevin time $t$ tending to infinity yield
the values of physical correlators without any assumptions about the model
of the vacuum.  In [2] the minimal closed set of such equations,
corresponding to the Gaussian distribution of fields, was obtained for the
case of gluodynamics and investigated in the lowest orders of perturbation
theory both in standard, non--gauge--invariant, and gauge--invariant ways.

The next related problem, one needs to solve in the framework of this
approach, is the problem of separation of perturbative gluonic contributions
in every term of cumulant expansion. To this end we split the total gluonic
field into a background and a quantum fluctuation:

\be
A^a_{\mu} = B^a_{\mu} + g Q^a_{\mu},
\ee
where the principle of this division is unimportant [6]. In particular, the
background field may be pure classical, and in this case we come to the
problem of quantization of classical solutions, but, generally speaking, the
fields $B^a_{\mu}$  form a quantum ensemble.

Such a separation was used in [6--8] in order to develop perturbative
theory in the confining QCD vacuum, which ensures the area law of an
averaged Wilson loop with the value of the string tension, known
phenomenologically, $\sigma \simeq 0.2 GeV^2$ [9]. In [7] this formalism was
applied to the case of QCD at finite temperatures.

It was shown in [6], that the confining background kills all the infra--red
singularities, the lowest gluon and ghost corrections to the charge
renorma-\\
lization were calculated, and it was found out, that the usual
logarithmic growth of $\alpha_s(R) $ in the empty space at large distances
in the one--loop approximation disappears in the presense of a background.
Instead of that, one obtains, that $\alpha_s(R)$ is saturated at the scale
of the inverse excitation mass of the transverse string vibration $m^2 \sim
2\pi \sigma \sim 1 GeV^2$.

In [8] unitary background gauges, where ghosts are either absent or
nonpropagating, were found, which  may help to describe hybrid states in
terms of physical polarizations of $Q^a_{\mu}$ only.

However, in all the papers [6--8] the nonperturbative background  fields
were considered as given, and this input was  parametrized, e.g. by the full
set of cumulants. The main goal of this paper is to derive equations,
starting from the Lagrangian, from which the correlators  of background
fields and of  the quantum fluctuations may be obtained simultaneously. It
seems na-\\
tural to apply for this purpose stochastic quantization method
[1], since the background field formalism [10,6], developed  within this
type of quantization, possesses no-ghost  property [11] as well as
stochastic quantization of gluodynamics in the empty space [12]. The point
is that all the individual stochastic diagrams, contributing to some
gauge--invariant quantity may  remain finite for $t \rightarrow +\infty$.
The main idea [13] is to transform the  gauge field in such a way, that in
the Langevin equation for the new field the projector onto transverse
degrees of freedom of gluonic field, standing in the action, is replaced by
an invertible matrix. This transformation is required to leave unchanged all
the gauge--invariant quantities, and, hence, should be a gauge
transformation, but depending on $t$ (because $t$--independent gauge
transformations leave the form of the Langevin equation invariant).

It is known [12], that it is not necessary to add a gauge--fixing term into
the Langevin equation, since the direct  iteration in powers of coupling
constant without introducing ghost fields leads to the same results as
Faddeev--Popov perturbation theory, because the Langevin time takes the role
of a gauge parameter. For example, the coefficient at the projector onto
longitudinal degrees of freedom of the gauge field in the free propagator is
lineary divergent at $t$ tending to infinity, but, if one fixes $t$,
calculates gauge--invariant quantities and then goes to the physical limit,
$t \rightarrow +\infty$, the divergent terms will cancel each other in the
same manner as the terms, depending on the gauge parameter in the framework
of the usual approach. That is why, in contrast to the Faddeev--Popov method
of quantization, which reproduces correctly only small field fluctuations
(since for the case of strong fields Gribov  ambiguities arise [14]),
Langevin equation does not distort nonperturbative effects.

Furthermore, it turns out, that the properly chosen $t$--dependent gauge
transformation modifies Langevin equation in such a way, that all the lineary
divergent terms disappear completely, which is useful for calculations
[13], and, in particular, suggests a method of quantization of
non--holonomic systems  [15]. This so--called
stochastic or Zwanziger gauge fixing procedure was applied in [11] to
quantization of gluodynamics in a background, and the  $\beta$--function in
the one--loop approximation was computed. In what follows we shall exploit
the Zwanziger term, introduced in the paper [11], to fix the gauge of a
quantum fluctuation, but, in contrast to [11], we shall not split Langevin
equation into two parts in the  sense of loop expansion, but use the total
one in order to derive equations for correlators both of a background and of
quantum fluctuations. This work is performed in section 2.

In section 3 we present three approaches to stochastic regularization of the
obtained equations. The first two covariant  derivative approaches use the
methods, suggested in the papers [16--19], while the alternative to them,
but also Markovian, third approach is a new one. It is then shown, that all
the three types of regularization leads to the properly regularized
expressions for the propagator of a background field gluon in the lowest
order of perturbation theory, when one neglects  perturbative gluonic
interactions.  The main results of the paper and possible future
developments are discussed in the Conclusion.

\vspace{3mm}
\section{Equations for correlators in bilocal approximation.}

\hspace{5mm} In this section we present a general method of derivation of an
infinite system of equations for correlators of  background fields
$B^a_{\mu}$, quantum fluctuations $Q^a_{\mu}$ and stochastic noise fields
$\eta^a_{\mu}$  and use it to obtain a minimal closed set of such equations,
corresponding to the so--called bilocal approximation, which follows from
the assumption, that the grand ensemble of fields is Gaussian, so that all
the  cumulants, higher than quadratic, are put equal to zero. Lattice data
suggest that this approximation has good accuracy in the confining regime of
an averaged Wilson loop (for a discussion see the last reference in [3]).
This hypothesis about the  predominancy of bilocal correlations in the
vacuum leads to the two alternative methods of investigation of higher
correlators:  the first one is based on the exact equations, where bilocal
and higher correlators are considered on the same footing, while the other
is the iterative one, where the values of bilocal correlators, obtained from
the minimal system of equations, are then used to calculate threelocal
correlators and so on. Moreover, in what follows we shall neglect all the
quantities higher than of the  second order in coupling constant, which
means, that the equations to be obtained will not reproduce correctly those
of Feynman diagrams, which contain three-- and four--perturbative--gluonic
vertices. In order to extract  explicitly the dependence on the coupling
constant, we shall deal below with the fields
$a^a_{\mu}=gA^a_{\mu},~b^a_{\mu}=gB^a_{\mu}$ and $q^a_{\mu}=g Q^a_{\mu}$.

A known important property of the background field method [10,6,8] is that
it is possible to introduce the gauge fixing term for $q^a_{\mu}$, which
breaks down the gauge invariance of the partition function under quantum
gauge transformations (which leave the background unchanged), but
preserves gauge invariance under the so--called background gauge
transformations

$$b_{\mu} \rightarrow U^+ (b_{\mu}+i\partial_{\mu})U,$$

\be
q_{\mu} \rightarrow U^+ q_{\mu}U.
\ee

This background gauge condition leads to the unique choice of Zwanziger
term, ensuring locality and renormalizability of the theory [11], so that
the Langevin equation takes the form

\be
\dot{a}^a_{\mu}=(D^{(a)}_{\lambda}{\cal{F}}^{(a)}_{\lambda\mu})^a +
g(D^{(a)}_{\mu} D^{(b)}_{\rho} q_{\rho})^a - g\eta^a_{\mu},
\ee
where

$${\cal{F}}^a_{\mu\nu}=\partial_{\mu}a^a_{\nu} -
\partial_{\nu}a^a_{\mu}+f^{abc}a^b_{\mu}a^c_{\nu},~~~~
(D^{(a)}_{\lambda}{\cal{F}}_{\lambda\mu})^a =
\partial_{\lambda}{\cal{F}}^a_{\lambda\mu} +
f^{abc}a^b_{\lambda}{\cal{F}}^c_{\lambda\mu},$$

\be
<\eta^a_{\mu}(x,t)\eta^b_{\nu}(x',t')> = 2\delta_{\mu\nu}
\delta^{ab}\delta(x-x')\delta(t-t'),
\ee
and the sign of $\eta^a_{\mu}$ is changed.

Due to (2), all the correlators, containing $q^a_{\mu}$ will be
gauge--invariant, while for the background field one should use Schwinger
gauge $b^a_{\mu}(x,t)(x-x_0)_{\mu}=0$ (where $x_0$ is an arbitrary point), in
which $b_{\mu}$ may be explicitly expressed through
${\cal{F}}^{(b)}_{\mu\nu}$:

$$b_{\mu}(x,t)=\int\limits^x_{x_0} dz_{\nu}\alpha
(z,x){\cal{F}}_{\nu\mu}(z,t),$$
where  $\alpha(z,x)\equiv \frac{(z-x_0)_{\nu}}{(x-x_0)_{\nu}}$, and here and
later in all the integrals of the type $\int\limits^x_{x_0}$ the path of
integration is a straight line. However, the final equations will be
gauge--invariant in the same way, as it was discussed in [2].

Introducing the generating functional

\be
\Phi_{\beta} = Pexp~i\oint\limits_C dx_{\mu} \left \{\int\limits^x_{x_0}
dz_{\nu}\alpha(z,x){\cal{F}}^{(b)}_{\nu\mu} (z,t)+g\Biggl (q_{\mu}(x,t) +
\beta \int\limits^t_0 dt'\eta_{\mu}(x,t') \Biggr ) \right \},
\ee
where $C$ is some
fixed closed contour and $\beta$ is a $c$--number, one obtains, using
Langevin equation (3):

$$tr\frac{\partial}{\partial t}<\Phi_{\beta}>=itr\oint\limits_C
dx_{\mu}<\Phi_{\beta}\Biggl(
D^{(a)}_{\lambda}({\cal{F}}^{(a)}_{\lambda\mu}(x,t)+
g\delta_{\lambda\mu}D^{(b)}_{\rho}q_{\rho}(x,t)) +$$

\be
+ (\beta-1)g\eta_{\mu}(x,t)
\Biggr)>.
\ee
Applying to (6) the formula [5]

\be
<e^AB> = <e^A> \Biggl ( <B> + \sum^{\infty}_{n=1} \frac{1}{n!}\ll A^nB \gg
\Biggr ),
\ee
where $A$ and $B$ are two arbitrary operators, we have in bilocal
approximation:

$$\frac{1}{2}tr \frac{\partial}{\partial t}
\ll V_{\nu}(y,x_0,t)V_{\mu}(u,x_0,t) \gg = tr \ll
V_{\nu}(y,x_0,t)\Biggl (
D^{(a)}_{\lambda}({\cal{F}}^{(a)}_{\lambda\mu}(u,x_0,t)+$$

\be
+ g \delta_{\lambda\mu}D^{(b)}_{\rho}q_{\rho}(u,x_0,t))\Biggr ) \gg,
\ee
where

$$V_{\mu}(y,x_0,t)=\int\limits^y_{x_0}dz_{\nu}\alpha(z,x)
{\cal{F}}^{(b)}_{\nu\mu}(z,x_0,t)+g\Biggl (
q_{\mu}(y,x_0,t)+\int\limits^t_0 dt' \eta_{\mu}(y,x_0,t,t')\Biggr ),$$

$${\cal{F}}^{(b)}_{\nu\mu}(z,x_0,t)=\Phi(x_0,z,t){\cal{F}}^{(b)}_{\nu\mu}(z,t)
\Phi(z,x_0,t), ~~~~~~q_{\mu}(y,x_0,t) =$$

$$= \Phi(x_0,y,t)q_{\mu}(y,t)\Phi(y,x_0,t), ~~~~\eta_{\mu}(y,x_0,t,t')=
\Phi(x_0,y,t) \eta_{\mu}(y,t')\Phi(y,x_0,t),$$

$$\Phi(z,x_0,t) = Pexp \Biggl [ i
\int\limits^z_{x_0}dz'_{\sigma}b_{\sigma}(z',t)\Biggr ].$$
Noticing, that, due to
(1),

$$D^{(a)}_{\lambda}=D^{(b)}_{\lambda}-ig [q_\lambda,\cdot
],~~~{\cal{F}}^{(a)}_{\lambda\mu}={\cal{F}}^{(b)}_{\lambda\mu}+g(D^{(b)}
_{\lambda}q_{\mu}-D^{(b)}_{\mu}q_{\lambda} - ig [q_{\lambda},q_{\mu}]),$$
one gets from (8) the first equation of bilocal approximation:

$$\frac{1}{2} tr \frac{\partial}{\partial
t}<V_{\nu}(y,x_0,t)V_{\mu}(u,x_0,t)> = tr\frac{\partial}{\partial
u_{\lambda}} \Biggl (
<V_{\nu}(y,x_0,t){\cal{F}}^{(b)}_{\lambda\mu}(u,x_0,t)> +$$

\be
+g\frac{\partial}{\partial
u_{\rho}}<V_{\nu}(y,x_0,t)G_{\rho\lambda\mu}(u,x_0,t)> \Biggr ),
\ee
where

$$G_{\rho\lambda\mu}(u,x_0,t)=\delta_{\rho\lambda}q_{\mu}(u,x_0,t)-
\delta_{\rho\mu}q_{\lambda}(u,x_0,t)+\delta_{\mu\lambda}
q_{\rho}(u,x_0,t),$$
and we put all the terms with space--time derivatives to the right hand
side. Here in order to disentangle the averages, containing covariant
derivatives, we used the formulae:

$$tr (D^{(b)}_{\mu}M(u,x_0)N)=tr\left\{ \frac{\partial}{\partial
u_{\mu}}M(u,x_0)N + i\int\limits^u_{x_0} dz_{\sigma}\alpha(z,u)\cdot
\right.$$

\be
\left. \cdot \Biggl (
M(u,x_0)N{\cal{F}}^{(b)}_{\mu\sigma}(z,x_0,t)-
M(u,x_0){\cal{F}}^{(b)}_{\mu\sigma}(z,x_0,t)N \Biggr )  \right \},
\ee

$$tr(D^{(b)}_{\mu}D^{(b)}_{\nu}M(u,x_0)N) = tr \left
\{\frac{\partial^2}{\partial u_{\mu}\partial u_{\nu}}
M(u,x_0)N+i\Biggl (M(u,x_0)N{\cal{F}}^{(b)}_{\nu\mu}(u,x_0,t)- \right.$$

$$-M(u,x_0){\cal{F}}^{(b)}_{\nu\mu}(u,x_0,t)N \Biggr
)+i\int\limits^u_{x_0}dz_{\sigma} \Biggl(
\alpha(z,u)\frac{\partial}{\partial u_{\nu}}\Biggl (M(u,x_0)
N{\cal{F}}^{(b)}_{\mu\sigma}(z,x_0,t)-$$

$$-M(u,x_0){\cal{F}}^{(b)}_{\mu\sigma}(z,x_0,t)N \Biggr ) +
\frac{\partial}{\partial u_{\mu}}\alpha(z,u)\Biggl (
M(u,x_0)N{\cal{F}}^{(b)}_{\nu\sigma}(z,x_0,t)-$$

$$-M(u,x_0){\cal{F}}^{(b)}_{\nu\sigma}(z,x_0,t)N \Biggr )\Biggr ) +
\int\limits^u_{x_0}dz_{\sigma}\alpha(z,u)\int\limits^u_{x_0}dw_{\zeta}
\alpha(w,u) \cdot $$

$$\cdot \Biggl (
M(u,x_0){\cal{F}}^{(b)}_{\mu\sigma}(z,x_0,t)N{\cal{F}}^{(b)}_{\nu\zeta}
(w,x_0,t)+M(u,x_0){\cal{F}}^{(b)}_{\nu\zeta}(w,x_0,t)N{\cal{F}}^{(b)}
_{\mu\sigma}(z,x_0,t)- $$

$$- M(u,x_0){\cal{F}}^{(b)}_{\nu\zeta}(w,x_0,t){\cal{F}}^{(b)}_{\mu\sigma}
(z,x_0,t)N -$$

\be
\left. -
M(u,x_0)N{\cal{F}}^{(b)}_{\mu\sigma}(z,x_0,t){\cal{F}}^{(b)}_{\nu\zeta}
(w,x_0,t) \Biggr ) \right \},
\ee
where $M(u, x_0)$ is equal to ${\cal{F}}^{(b)}_{\alpha\beta}$ $(u,x_0,t)$
or $q_{\alpha}(u,x_0,t),$ $N\equiv $\\
$\equiv N_{\mu_1...\mu_n}(x_1,t_1,...,x_n,t_n,x_0,t)$ is, generally
speaking, a product of some number of $F_{\alpha\beta},~q_{\alpha}$ and
$\eta_{\alpha}$, which are given in the points $x_1 \not= u,...,x_n \not=u$
at the moments $t_1,..., t_n$ of Langevin time respectively, where all the
parallel transporters between $x_0$ and each of these points are built of
the field $b^a_{\mu}$ and given at the same moment $t$.

Differentiating equation (6)  twice by $\beta$, putting then $\beta$ equal
to 1, using the formulae (7), (10) and (11) and the definitions of three--
and fourlocal path--ordered cumulants [4,2], one obtains two more equations
of bilocal approximation, where all the perturbative correlators, higher
than of the second order in coupling constant are neglected:

$$tr \left \{\frac{\partial}{\partial t}<
V_{\nu}(y,x_0,t)\eta_{\mu}(u,x_0,t,t')> - \right.$$

$$-\frac{1}{2}\oint\limits_C dv_{\xi}\oint\limits_C dw_{\sigma}
\Biggl ( \frac{\partial}{\partial t} <V_{\xi}(v,x_0,t)V_{\sigma}(w,x_0,t)>
\Biggr ) <V_{\nu}(y,x_0,t)\eta_{\mu}(u,x_0,t,t')> +$$

$$+g\int\limits^u_{x_0}
dz_{\sigma}\alpha(z,u)\int\limits^u_{x_0}dw_{\zeta}(w,u)\Biggl ( <
G_{\rho\lambda\mu}(u,x_0,t) {\cal{F}}^{(b)}_{\rho\zeta} (w,x_0,t)> \cdot $$

$$\cdot <{\cal{F}}^{(b)}_{\lambda\sigma}(z,x_0,t)\eta_{\nu}(y,x_0,t,t')>+ $$

$$+ < G_{\rho\lambda\mu}(u,x_0,t)\eta_{\nu}(y,x_0,t,t')><
{\cal{F}}^{(b)}_{\lambda\sigma}(z,x_0,t)
{\cal{F}}^{(b)}_{\rho\zeta}(w,x_0,t)>-$$

$$-<G_{\rho\lambda\mu}(u,x_0,t){\cal{F}}^{(b)}_{\lambda\sigma}(z,x_0,t) >
<\eta_{\nu}(y,x_0,t,t'){\cal{F}}^{(b)}_{\rho\zeta}(w,x_0,t)>-$$

$$-<G_{\rho\lambda\mu}(u,x_0,t) {\cal{F}}^{(b)}_{\rho\zeta}(w,x_0,t)>
<\eta_{\nu}(y,x_0,t,t'){\cal{F}}^{(b)}_{\lambda\sigma}(z,x_0,t) > \Biggr )
+$$

$$+ g^2\int\limits^u_{x_0}dz_{\sigma}\alpha(z,u) \Biggl ( <
G_{\rho\lambda\mu} (u,x_0,t) {\cal{F}}^{(b)}_{\rho\sigma}(z,x_0,t)>
<\eta_{\nu}(y,x_0,t,t')q_{\lambda}(u,x_0,t)> + $$

$$+<G_{\rho\lambda\mu}(u,x_0,t)q_{\lambda} (u,x_0,t)><\eta_{\nu}(y,x_0,t,t')
{\cal{F}}^{(b)}_{\rho\sigma}(z,x_0,t)> - $$

$$-<G_{\rho\lambda\mu}(u,x_0,t) \eta_{\nu}(y,x_0,t,t')><q_{\lambda} (u,x_0,t)
{\cal{F}}^{(b)}_{\rho\sigma}(z,x_0,t)> - $$

$$\left.-<G_{\rho\lambda\mu} (u,x_0,t) {\cal{F}}^{(b)}_{\rho\sigma}(z,x_0,t)>
<q_{\lambda} (u,x_0,t) \eta_{\nu}(y,x_0,t,t')> \Biggr ) \right \} = $$

$$= tr \left \{ g\frac{\partial^2}{\partial u_{\lambda}\partial
u_{\lambda}} < q_{\mu}(u,x_0,t) \eta_{\nu}(y,x_0,t,t')>+ \right.$$

\be
\left.+\frac{\partial}{\partial u_{\lambda}}<{\cal{F}}^{(b)}_{\lambda\mu}
(u,x_0,t)\eta_{\nu}(y,x_0,t,t')> \right \}.
\ee

$$tr \left \{ \int\limits^t_0 dt''\frac{\partial}{\partial t}
<\eta_{\nu}(y,x_0,t,t'')\eta_{\mu}(u,x_0,t,t')>+<
\eta_{\nu}(y,x_0,t,t)\eta_{\mu}(u,x_0,t,t')>-
\right.$$

$$-<\eta_{\nu}(y,x_0,t,t')\eta_{\mu}(u,x_0,t,t)> -$$

$$ - \frac{1}{2}\oint\limits_C
dv_{\xi} \oint\limits_C dw_{\sigma}\Biggl ( \frac{\partial}{\partial t}
<V_{\xi}(v,x_0,t)V_{\sigma}(w,x_0,t)> \Biggr ) \cdot$$

$$\cdot \int\limits^t_0 dt''<\eta_{\nu}(y,x_0,t,t'')\eta_{\mu}(u,x_0,t,t')>
+\oint\limits_C dz_{\sigma} \int\limits^t_0 dt'' \Biggl ( \int\limits^u_{x_0}
dw_{\zeta}\alpha(w,u)\cdot$$

$$\cdot\Biggl ( <{\cal{F}}^{(b)}_{\lambda\mu}(u,x_0,t)
\eta_{\sigma}(z,x_0,t,t'')><\eta_{\nu}(y,x_0,t,t')
{\cal{F}}^{(b)}_{\lambda\zeta}(w,x_0,t)>-$$

$$-<{\cal{F}}^{(b)}_{\lambda\mu}(u,x_0,t){\cal{F}}^{(b)}_{\lambda\zeta}(w,x_0,t)>
<\eta_{\sigma}(z,x_0,t,t'')\eta_{\nu}(y,x_0,t,t')> \Biggr ) +$$

$$+g \Biggl ( <G_{\rho\lambda\mu}(u,x_0,t) \eta_{\sigma}(z,x_0,t,t'')>
<\eta_{\nu}(y,x_0,t,t')
{\cal{F}}^{(b)}_{\rho\lambda}(u,x_0,t)>+ $$

$$+2 <q_{\lambda}(u,x_0,t){\cal{F}}^{(b)}_{\mu\lambda}(u,x_0,t)>
<\eta_{\sigma}(z,x_0,t,t'')\eta_{\nu}(y,x_0,t,t')> \Biggr )\Biggr )=$$

$$=g\oint\limits_C dz_{\sigma}\int\limits^t_0 dt'' \Biggl (
\int\limits^u_{x_0} dw_{\zeta}\Biggl ( \alpha(w,u)
\frac{\partial}{\partial u_{\rho}} \Biggl ( <
G_{\rho\lambda\mu}(u,x_0,t){\cal{F}}^{(b)}_{\lambda\zeta}(w,x_0,t)> \cdot$$

$$\cdot <\eta_{\sigma}(z,x_0,t,t'')\eta_{\nu}(y,x_0,t,t')>-
<G_{\rho\lambda\mu}(u,x_0,t)\eta_{\sigma}(z,x_0,t,t'')> \cdot $$

$$\cdot <\eta_{\nu}(y,x_0,t,t'){\cal{F}}^{(b)}_{\lambda\zeta}(w,x_0,t)>
\Biggr ) +\frac{\partial}{\partial u_{\lambda}}\alpha(w,u) \Biggl (
<G_{\rho\lambda\mu}(u,x_0,t) \cdot $$

$$\cdot {\cal{F}}^{(b)}_{\rho\zeta}(w,x_0,t)>
<\eta_{\sigma}(z,x_0,t,t'')
\eta_{\nu}(y,x_0,t,t')> - $$

$$- < G_{\rho\lambda\mu}(u,x_0,t)
\eta_{\sigma}(z,x_0,t,t'')>
<\eta_{\nu}(y,x_0,t,t') {\cal{F}}^{(b)}_{\rho\zeta}(w,x_0,t)>
\Biggr ) \Biggr ) + $$

$$ + g\Biggl (  \frac{\partial}{\partial u_{\rho}}
<G_{\rho\lambda\mu}(u,x_0,t)\eta_{\sigma}(z,x_0,t,t'')> \Biggr )
<\eta_{\nu}(y,x_0,t,t') q_{\lambda}(u,x_0,t)> - $$

\be
- g<q_{\lambda}(u,x_0,t)
\eta_{\sigma}(z,x_0,t,t'')>
\frac{\partial}{\partial u_{\rho}} <\eta_{\nu}(y,x_0,t,t')
G_{\rho\lambda\mu}(u,x_0,t)> \Biggr ).
\ee

As was discussed in [2], in the physical limit, $t \rightarrow +\infty$, in
the confi-\\
ning regime of an averaged Wilson loop, the dependence on the
point $x_0$ is negligible, since the difference between each of the
cumulants and its gauge--invariant analog is of the order of
 $\frac{T^2_g}{R^2}\leq
0.04$ [2], where $T_g$ is the correlation length of the vacuum, at which a
cumulant vanishes and $R$ is a space width of a Wilson loop, and we obtain
gauge--invariant equations for correlators of the fields ${\cal{F}}
^{(b)}_{\mu\nu}$ and  $q_\mu$, $\eta_{\mu}$, immersed into a background
 as insertions in the parallel transporters.

Note, that among these equations only the first one, equation (9), is linear,
while equations (12) and (13) produce complicated hierarchy of perturbative
correlators up to the second order of perturbation theory. Hence,  the procedure of solution
of equations (9),(12) and (13) is the following: first one needs to put all the
perturbative fluctuations equal to zero and to solve the so--reduced equations
for correlators of a background and of the Gaussian noise fields (which are just
the equations (17),(19) and (20) from the paper [2]). After that one should to
include perturbative interactions, expanding the correlators, containing
$q_{\mu}$ up to the order of $g^2$  and using the obtained values of pure background
and noise fields$^{\prime}$ correlators.

\vspace{3mm}
\section{Stochastic regularization and perturbative expansion of equations
(9),(12) and (13).}

\hspace{5mm} In this section we present three methods of stochastic regularization
of equations (9),(12) and (13) and use them to derive regularized propagators of a background
in the lowest order of perturbation theory in the absence of perturbative
corrections. The first two of them are based on covariant derivative
regularization schemes, which were suggested in the papers [16--18] and used
in [19] to calculate $\beta$--function in QCD in the one--loop
approximation.

First is the so--called power--law regularization scheme

$$\eta^a_{\mu}(x,t) \rightarrow \int dy R^{ab}(x,y,t) \eta^b_{\mu}(y,t),$$
where

$$R^{ab}(x,y,t) = \Biggl [ \frac{1}{(1-\frac{\Delta}{\Lambda^2})^n} \Biggr ]^
{ab} (x,y,t), ~~~~n=1,2, ...,$$
$\Lambda$ is an ultraviolet cutoff, $\Delta^{ab}(x,y,t) \equiv \int
dz(D_{\mu})^{ac}(x,z,t)(D_{\mu})^{cb}(z,y,t)$ is the covariant Laplacian with
$(D_{\mu})^{ab}(x,y,t)\equiv D^{(b)ab}_{\mu} (x,t)\delta(x-y)$. One of the
main results of the paper [17] is the proof, that any Yang--Mills theory in $d$
dimensions is regularized to all the orders, when we  choose $n\geq
[\frac{1}{2} (d+1)]$, where $[x]$ is the largest integer less than or equal
to $x$.

The generalization of equations (9), (12) and (13) after applying such a regularization
is obvious. For example, equation (9), written in details, takes the form:

$$\frac{1}{2}tr\frac{\partial}{\partial t}\left\{
\int\limits^y_{x_0}dz_{\lambda}(z,y)\int\limits^u_{x_0}dx_{\rho}\alpha(x,u)
<{\cal{F}}^{(b)}_{\lambda\nu}(z,x_0,t){\cal{F}}^{(b)}_{\rho\mu}(x,x_0,t)>
+\right.$$

$$+g\int\limits^y_{x_0}dz_{\lambda}\alpha(z,y)\Biggl
(<{\cal{F}}^{(b)}_{\lambda\nu}(z,x_0,t) q_{\mu}(u,x_0,t)>+$$

$$+\int\limits^t_0 dt'\int dw<{\cal{F}}^{(b)}_{\lambda\nu}(z,x_0,t)
\xi_{\mu}(u,w,x_0,t,t')>\Biggr )+g\int\limits^u_{x_0}dx_{\rho}\alpha(x,u)
\Biggl (<q_{\nu}(y,x_0,t)\cdot $$

$$\cdot{\cal{F}}^{(b)}_{\rho\mu}(x,x_0,t)>+\int\limits^t_0 dt'\int dw
<\xi_{\nu}(y,w,x_0,t,t'){\cal{F}}^{(b)}_{\rho\mu}(x,x_0,t)> \Biggr )+$$

$$+g^2<q_{\nu}(y,x_0,t)q_{\mu}(u,x_0,t)>+g^2\int\limits^t_0 dt'\int dw
\cdot $$

$$ \cdot \Biggl (<q_{\nu}(y,x_0,t)\xi_{\mu}(u,w,x_0,t,t')>
+<\xi_{\nu}(y,w,x_0,t,t')q_{\mu}(u,x_0,t)>+$$

$$\left.+\int\limits^t_0 dt''\int dv <\xi_{\nu}(y,w,x_0,t,t')
\xi_{\mu}(u,v,x_0,t,t'')>\Biggr )\right \}=$$

$$=tr\frac{\partial}{\partial u_{\lambda}} \left \{ \int\limits^y_{x_0}
dz_{\sigma}\alpha(z,y)\Biggl (<{\cal{F}}^{(b)}_{\sigma\nu}(z,x_0,t)
{\cal{F}}^{(b)}_{\lambda\mu}(u,x_0,t)>+ \right.$$

$$+g\frac{\partial}{\partial
u_{\lambda}}<{\cal{F}}^{(b)}_{\sigma\nu}(z,x_0,t)
q_{\mu}(u,x_0,t)>-g\frac{\partial}{\partial u_{\mu}}
<{\cal{F}}^{(b)}_{\sigma\nu}(z,x_0,t)q_{\lambda}(u,x_0,t)>+$$

$$+g\delta_{\lambda\mu}\frac{\partial}{\partial u_{\rho}}
<{\cal{F}}^{(b)}_{\sigma\nu}(z,x_0,t)q_{\rho}(u,x_0,t)>\Biggr ) +
g<q_{\nu}(y,x_0,t){\cal{F}}^{(b)}_{\lambda\mu}(u,x_0,t)>+ $$

$$+g\int\limits^t_0 dt'\int dw
<\xi_{\nu}(y,w,x_0,t,t'){\cal{F}}^{(b)}_{\lambda\mu}(u,x_0,t)>+$$

$$+g^2\Biggl ( \frac{\partial}{\partial u_{\lambda}}<q_{\nu}(y,x_0,t)
q_{\mu}(u,x_0,t)> - \frac{\partial}{\partial u_{\mu}} <q_{\nu}
(y,x_0,t)q_{\lambda}(u,x_0,t)> +$$

$$+\delta_{\lambda\mu}\frac{\partial}{\partial u_{\rho}}
<q_{\nu}(y,x_0,t)q_{\rho}(u,x_0,t)> + $$

$$+\int\limits^t_0 dt' \int dw \Biggl ( \frac{\partial}{\partial u_{\lambda}}
<\xi_{\nu}(y,w,x_0,t,t') q_{\mu}(u,x_0,t)>-$$

$$-\frac{\partial}{\partial u_{\mu}}
<\xi_{\nu}(y,w,x_0,t,t') q_{\lambda}(u,x_0,t)> + $$

$$\left.+ \delta_{\lambda\mu}\frac{\partial}{\partial u_{\rho}}
<\xi_{\nu}(y,w,x_0,t,t') q_{\rho}(u,x_0,t)> \Biggr ) \Biggr ) \right \}, $$
where $\xi_{\mu}(u,w,x_0,t,t') \equiv \Phi(x_0,u,t)\Biggl
[\frac{1}{(1-\frac{\Delta}{\Lambda^2})^n}\Biggr ]^{ab} (u,w,t')\eta^b_{\mu}
(w,t')t^a\Phi(w,x_0,t)$.
The equations (12) and (13) changes correspondingly.

In the lowest order of perturbation theory, when one neglects perturbative
fluctuations, and equations (9), (12) and (13) reduce to the equations for
background fields only, one may check in the same way, as it was done in [2]
for the unregularized case, that for $n=2(d=4)$ the regularized propagator,
obtained from these equations, has the form:

$$ <B^a_{\mu}(x,t) B^b_{\nu}(y,t)>=\delta^{ab}\int\frac{dk}{(2\pi)^4}
e^{-ik(x-y)} \frac{(\Lambda^2)^4}{k^2(k^2+\Lambda^2)^4}\Biggl (
(1-e^{-2k^2t})T_{\mu\nu}+ $$

\be
+ 2k^2 tL_{\mu\nu} \Biggr ),
\ee
where $T_{\mu\nu}\equiv\delta_{\mu\nu}-\frac{k_{\mu}k_{\nu}}{k^2},
L_{\mu\nu}\equiv\frac{k_{\mu}k_{\nu}}{k^2}$ are the transverse and the
longitudinal projectors respectively. In (14) one may recognize the
expression (2.19) from the paper [17], where the "gauge--fixing parameter"
$\alpha=2 k^2t$, which seems to be natural in the sense, stated in the
Introduction.

Second method of regularization exploits the so--called heat--kernel
re-\\
gularization scheme [18], where the regulator has the form
$R^{ab}(x,y,t)= \\
=(e^{\frac{\Delta}{\Lambda^2}})^{ab}(x,y,t)$, that is
believed to be technically superior for nonperturbative analysis. In
analogous way, one gets from the equations (9) and (12) in the lowest order,
when the perturbative gluons$^{\prime}$ contributions are neglected:

\be
<B^a_{\mu}(x,t)B^b_{\nu}(y,t) > = \delta^{ab} \int\frac{dk}{(2\pi)^4} e^{-ik(x-y)}
\frac{e^{-\frac{2k^2}{\Lambda^2}}}{k^2}
\Biggl ( (1-e^{-2k^2t})T_{\mu\nu} + 2k^2t L_{\mu\nu} \Biggr ),
\ee
which coincides with the formula (9) from the paper [18] at $\alpha=2k^2t$.

Finally, let us present a new one, also Markovian and preserving gauge
invariance, type of regularization \footnote{This method was suggested by
Professor Yu.A.Simonov.}. Its basic idea is to smear space--time
delta--function in (4) so, that properly modified Langevin equation remains
gauge--invariant. To this end we introduce the new noise fields

\be
\xi^a_{\mu}(x,t) \equiv \frac{(\Lambda^2)^2}{4\pi^2} \int dy~e^{-\frac{\Lambda^2
(x-y)^2}{2}} \eta^a_{\mu} (y,t),
\ee
so that

\be
<\xi^a_{\mu}(x,t) \xi^b_{\nu}(y,t') > = \frac{(\Lambda^2)^2}{8\pi^2} \delta_{\mu
\nu} \delta^{ab} \delta(t-t') e^{-\frac{\Lambda^2(x-y)^2}{4}},
\ee

$$\lim\limits_{\mid \Lambda\mid \rightarrow +\infty}<\xi^a_{\mu}(x,t)
\xi^b_{\nu}(y,t')>=
2\delta_{\mu\nu}\delta^{ab} \delta(x-y)\delta(t-t')$$

and modify the Langevin equation (3) in gauge--invariant way

$$
\dot{a}^a_{\mu}=\Biggl(D^{(a)}_{\lambda}{\cal{F}}^{(a)}_{\lambda\mu}\Biggr )
^a + g\Biggl (D^{(a)}_{\mu}D^{(b)}_{\rho}q_{\rho}\Biggr )^a - $$

 \be
- \frac{g(\Lambda^2)^2}{4\pi^2}\int dy~e^{-\frac{\Lambda^2(x-y)^2}{2}}
\Phi(x,y,t)\eta_{\mu}(y,t)\Phi(y,x,t),
\ee
keeping in mind, that at $\mid\Lambda\mid \rightarrow +\infty$ the integral
in the right hand side is saturated at $\mid y-x\mid \ll
\frac{1}{\mid\Lambda\mid}$.

Then, due to (16) and (17), in the lowest order of perturbation theory in the
absence of perturbative gluons, it follows from the equations (9) and (12),
regularized according to (18), correspondingly

$$\frac{1}{2}\frac{\partial}{\partial t}\Biggl (<B^a_{\nu}(y,t)B^b_{\mu}
(u,t)>+\frac{(\Lambda^2)^2}{4\pi^2}\int\limits^t_0 dt'\int dv \Biggl (
e^{-\frac{\Lambda^2(u-v)^2}{2}}<B^a_{\nu}(y,t)\eta^b_{\mu}(v,t')>+$$

$$+e^{-\frac{\Lambda^2(y-v)^2}{2}}<B^a_{\mu}(u,t)\eta^b_{\nu}(v,t')>
\Biggr ) \Biggr ) + \frac{(\Lambda^2)^4}{32\pi^4}\int\limits^t_0 dt'
\int dv dv' e^{-\frac{\Lambda^2((y-v)^2+(u-v')^2)}{2}}\cdot $$

$$\cdot \Biggl (<\eta^a_{\nu}(v,t)\eta^b_{\mu}(v',t')>+<\eta^a_{\nu}(v,t'
)\eta^b_{\mu}(v',t)> \Biggr )=\Biggl (
\frac{\partial^2}{\partial
u_{\rho}\partial u_{\rho}}\delta_{\mu\lambda}-
\frac{\partial^2}{\partial
u_{\mu}\partial u_{\lambda}}\Biggr ) \cdot$$

\be
\cdot \Biggl ( <B^a_{\lambda}(u,t)B^b_{\nu}(y,t)> +
\frac{(\Lambda^2)^2}{4\pi^2}\int\limits^t_0 dt'\int dv
e^{-\frac{\Lambda^2(y-v)^2}{2}}<B^a_{\lambda}(u,t)\eta^b_{\nu}(v,t')>\Biggr
),
\ee

$$\int dv e^{-\frac{\Lambda^2(u-v)^2}{2}}\frac{\partial}{\partial t}
<B^a_{\nu}(y,t)\eta^b_{\mu}(v,t')>+\Biggl (
\frac{\partial^2}{\partial
u_{\mu}\partial u_{\lambda}} -
\frac{\partial^2}{\partial
u_{\rho}\partial u_{\rho}}\delta_{\mu\lambda}\Biggr ) \cdot$$

\be
\cdot\int dve^{-\frac{\Lambda^2(y-v)^2}{2}}
<B^a_{\lambda}(u,t)\eta^b_{\nu}(v,t')>=-\frac{1}{2}\delta_{\mu\nu}\delta^{ab}
\delta(t-t') e^{-\frac{\Lambda^2(y-u)^2}{4}}.
\ee

Looking for $<B^a_{\nu}(y,t)\eta^b_{\mu}(u,t')>$ in the form
$\delta^{ab}d_{\mu\nu}(z,\tau)$, where $z=u- -y,\tau=\mid t-t'\mid,
d_{\nu\mu}(z,\tau)=d_{\mu\nu}(z,\tau),~
d_{\mu\nu}(-z,\tau)=d_{\mu\nu}(z,\tau)$, one obtains from (20):

\be
\bar{d}_{\mu\nu}(k,\tau)=-\frac{8\pi^2}{(\Lambda^2)^2}\theta(\tau)
e^{-\frac{k^2}{\Lambda^2}}(T_{\mu\nu}e^{-k^2\tau}+L_{\mu\nu}) ,
\ee
where $\bar{d}_{\mu\nu}(x,\tau)\equiv \int dy
e^{-\frac{\Lambda^2(x+y)^2}{2}} d_{\mu\nu}(y,\tau)$.

Looking for $<B^a_{\nu}(y,t)B^b_{\mu}(u,t)>$ in the form
$\delta^{ab}h_{\mu\nu}(z,t)$, where $h_{\nu\mu}(z,t)=$\\
$=h_{\mu\nu}(z,t),
h_{\mu\nu}(-z,t)=h_{\mu\nu}(z,t)$ and using (4) and (20), one gets from the
equation (19):

$$\Biggl ( \frac{1}{2}\delta_{\mu\lambda}\frac{\partial}{\partial t}+
\frac{\partial}{\partial z_{\mu}\partial z_{\lambda}}-
\frac{\partial}{\partial z_{\rho}\partial z_{\rho}}\delta_{\mu\lambda}\Biggr
) h_{\lambda\nu}(z,t)= -\frac{(\Lambda^2)^2}{4\pi^2}\bar{d}_{\mu\nu}(z,0),$$
and, hence, due to (21), the regularized propagator of a background field
has the form

\be
<B^a_{\mu}(x,t)B^b_{\nu}(y,t)>=\delta^{ab}\int\frac{dk}{(2\pi)^4}e^{-ik(x-y)}
\frac{e^{-\frac{k^2}{\Lambda^2}}}{k^2} \Biggl (
(1-e^{-2k^2t})T_{\mu\nu}+2k^2tL_{\mu\nu}\Biggr ).
\ee
Therefore, we see, that this method of regularization is similar to the
heat--kernel method, but is simpler, since there do not arise higher
derivatives in  the regulator.

\section{Conclusion}
\hspace{5mm} In this paper we applied stochastic quantization [1] to develop
a method of derivation of an infinite system of exact equations for
gauge--invariant correlators in gluodynamics, where all the perturbative contributions are extracted
explicitly in the form of insertions into background parallel transporters.
Therefore, the obtained equations allow one to derive pure background  correlators
and the  correlators, containing perturbative corrections, simultaneously,
using for quantization the same stochastic noise fields.

After that we obtained the minimal set of equations of bilocal approximation
(corresponding to the Gaussian distribution of fields), where we threw
away all the perturbative interactions higher than of the second order, and
suggested for it three methods of stochastic regularization, all of which
preserve gauge invariance of the obtained equations. The first two of them,
based on the so--called covariant derivative regularization schemes, lead,
in the lowest order of perturbative theory in the absence of perturbative
gluons, to the known values of the regularized propagator of a background
field, while the third method is a new one. It yields the results, similar
to the heat--kernel method, but is simpler than the latter, since in the
framework of this method higher derivatives in the regulator do not exist.

The application of the suggested approach to treating the large--$N$ regime
of QCD as well as the new equation for the Master field  and its connection
with the Bootstrap equation will be a topic of a separate publication.
Possible types of solutions of the derived equations will be presented
elsewhere.

The results, presented in this paper, were partially reported
at the International Workshop "Nonperturbative Approaches to QCD", Trento,
Italy, July 10--29, 1995.

\vspace{3mm}
\section{Acknowledgements}

\hspace{5mm} I am grateful to Professor Yu.A.Simonov for useful discussions
and to M.Markina for typing the manuscript.

The work is supported by the Russian Fundamental Research Foundation, Grant
No.93--02--14937 and by the INTAS, Grant No.94--2851.

\newpage
{\Large \bf References}

\vspace{5mm}
\noindent
1. {\it Parisi G.} and {\it Wu Y.}, Scienta Sinica 1981, vol.24, p.483;
for a review see

{\it Damgaard P.H.} and {\it H\"uffel H.}, Phys.Rep. 1987,
vol.152, pp.227--398.

\noindent
2. {\it Antonov D.V.} and {\it Simonov Yu.A},  International Journal

of Modern Physics A (in press).

\noindent
3. {\it Dosch H.G.},  Phys.Lett.B, 1987, vol.190, p.177; {\it Simonov
Yu.A.},

 Nucl.Phys.B, 1988, vol.307, p.512; {\it Dosch H.G.} and {\it Simonov Yu.A.},

Phys.Lett.B, 1988, vol.205, p.339, Z.Phys.C, 1989, vol.45, p.147;

{\it Simonov Yu.A.}, Nucl.Phys.B, 1989, vol.324, 67, Phys.Lett.B, 1989,

vol.226, p.151, Phys.Lett.B, 1989, vol.228, p.413; for a review see

{\it Simonov Yu.A.}, Yad.Fiz., 1991, vol.54, p.192.

\noindent
4. {\it Simonov Yu.A.}, Yad.Fiz., 1989, vol.50, p.213.

\noindent
5. {\it Van Kampen N.G.}, Stochastic processes in physics and chemistry,

North--Holland Physics Publishing, 1984.

\noindent
6. {\it Simonov Yu.A.}, HD--THEP--93--16.

\noindent
7. {\it Simonov Yu.A.}, Yad.Fiz., 1995, vol.58, p.357; {\it Gubankova E.L.}

and {\it Simonov Yu.A.}. Phys.Lett.B, 1995, vol.360, p.93.

\noindent
8. {\it Dubin A.} and {\it Wyler D.}, HEP--TH/9512037 (Submit. to
Nucl.Phys.B).

\noindent
9. {\it Campostrini M. et al.}, Z.Phys.C, 1984, vol.25, p.173; {\it
Di Giacomo A.}

 and {\it Panagopoulos H.}, Phys.Lett.B, 1992, vol.285, p.133.

\noindent
10. {\it De Witt B.S.}, Phys.Rev., 1967, vol.162, pp.1195,1239; {\it
Honerkamp J.},

Nucl.Phys.B, 1972, vol.48, p.269; {\it t'Hooft G.}, Nucl.Phys., 1973, vol.62,

p.444; {\it Abbot L.F.}, Nucl.Phys., 1981, vol.185, p.189.

\noindent
11. {\it Okano K.}, Nucl.Phys.B, 1987, vol.289, p.109.

\noindent
12. {\it Namiki M. et al.}, Progr.Theor.Phys., 1983, vol.69, p.1580.

\noindent
13. {\it Zwanziger D.}, Nucl.Phys.B, 1981, vol.192, p.259, Nucl.Phys.B,

1982, vol.209, p.336; {\it Baulieu L.} and {\it Zwanziger D.}, Nucl.Phys.B,

1981, vol.193, p.163; {\it Horibe M. et al.}, Progr.Theor.Phys., 1983,
vol.70,

p.1636; {\it Sakamoto J.}, Progr.Theor.Phys., 1984, vol.71, p.881;

{\it Floratos E. et.al.},
Nucl.Phys.B., 1984, vol.241, p.221, Nucl.Phys.B, 1983,

vol.214, p.392; {\it Baulieu L.}, Phys.Lett.B, 1986, vol.171, p.396,

Nucl.Phys.B, 1986, vol.270, p.507; {\it Chan M.S.} and {\it Halpern M.B.},

Phys.Rev.D, 1986, vol.33,p.540.

\noindent
14. {\it Gribov V.N.}, Nucl.Phys.B, 1978, vol.139, p.1; {\it Horibe M. et
al.},

Progr.Theor.Phys., 1983, vol.70, p.1636.

\noindent
15. {\it Nakagoshi H. et al.}, Progr.Theor.Phys., 1983, vol.70, p.326.

\noindent
16. {\it Bern Z. et al.}, Phys.Lett.B, 1985, vol.165, p.151.

\noindent
17. {\it Bern Z. et al.}, Nucl.Phys.B, 1987, vol.284, p.35.

\noindent
18. {\it Bern Z. et al.}, Phys.Rev.D., 1987, vol.35, p.753.

\noindent
19. {\it Bern Z. et al.}, Nucl.Phys.B, 1987, vol.284, p.92.
\end{document}